# Spatial coherence of room-temperature monolayer WSe$_2$ exciton-polaritons in a trap

Hangyong Shan[1,†], Lukas Lackner[1], Bo Han[1], Evgeny Sedov[2,3,4], Christoph Rupprecht[5], Heiko Knopf[6,7,8], Falk Eilenberger[6,7,8], Johannes Beierlein[5], Nils Kunte[1], Martin Esmann[1], Kentaro Yumigeta[9], Kenji Watanabe[10], Takashi Taniguchi[11], Sebastian Klembt[5], Sven Höfling[5], Alexey V. Kavokin[2,3,12], Sefaattin Tongay[9], Christian Schneider[1,†] and Carlos Anton-Solanas[1,†]

[1]Institute of Physics, Carl von Ossietzky University, 26129 Oldenburg, Germany.
[2]School of Science, Westlake University, 18 Shilongshan Road, Hangzhou 310024, Zhejiang Province, People's Republic of China
[3]Institute of Natural Sciences, Westlake Institute for Advanced Study, 18 Shilongshan Road, Hangzhou 310024, Zhejiang Province, People's Republic of China
[4]Vladimir State University named after A. G. and N. G. Stoletovs, Gorky str. 87, 600000, Vladimir, Russia
[5]Technische Physik, Universität Würzburg, D-97074 Würzburg, Am Hubland, Germany.
[6]Institute of Applied Physics, Abbe Center of Photonics, Friedrich Schiller University, 07745 Jena, Germany.
[7]Fraunhofer-Institute for Applied Optics and Precision Engineering IOF, 07745 Jena, Germany.
[8]Max Planck School of Photonics, 07745 Jena, Germany.
[9]School for Engineering of Matter, Transport, and Energy, Arizona State University, Tempe, Arizona 85287, USA
[10]Research Center for Functional Materials, National Institute for Materials Science, 1-1 Namiki, Tsukuba 305-0044, Japan
[11]International Center for Materials Nanoarchitectonics, National Institute for Materials Science, 1-1 Namiki, Tsukuba 305-0044, Japan
[12]Russian Quantum Center, Skolkovo IC, Bolshoy Boulevard 30, bld. 1, 121205, Moscow, Russia
†Corresponding author. Email: hangyong.shan@uni-oldenburg.de, christian.schneider@uni-oldenburg.de, carlos.anton-solanas@uni-oldenburg.de

## ABSTRACT

**The emergence of spatial and temporal coherence of light emitted from solid-state systems is a fundamental phenomenon, rooting in a plethora of microscopic processes. It is intrinsically aligned with the control of light-matter coupling, and canonical for laser oscillation. However, it also emerges in the superradiance of multiple, phase-locked emitters, and more recently, coherence and long-range order have been investigated in bosonic condensates of thermalized light, as well as in exciton-polaritons driven to a ground state via stimulated scattering.**

**Here, we experimentally show that the interaction between photons in a Fabry-Perot microcavity and excitons in an atomically thin WSe$_2$ layer is sufficient such that the system enters the hybridized regime of strong light-matter coupling at ambient conditions. Via Michelson interferometry, we capture clear evidence of increased spatial and temporal coherence of the emitted light from the spatially confined system ground-state. The coherence build-up is accompanied by a threshold-like behaviour of the emitted light intensity, which is a fingerprint of a polariton laser effect. Valley-physics is manifested in the presence of an external magnetic field, which allows us to manipulate K and K' polaritons via the Valley-Zeeman-effect. Our findings are of high application relevance, as they confirm the possibility to use atomically thin crystals as simple and versatile components of coherent light-sources, and in valleytronic applications at room temperature.**

## INTRODUCTION

Atomically thin transition metal dichalcogenide crystals have emerged as a highly interesting material class in opto-electronic [1,2] and nanophotonic[3] applications because of the giant light-matter coupling strength they exhibit. They are profusely investigated to provide efficient light-to-current converters of great flexibility[4,5], and are considered as gain-material in micro-and nanolasers[6,7]. However, the strong Coulomb-interactions, which dominate the optical response up to room temperature in those materials, and which make them particularly promising for fundamental studies of many-particle and quantum correlated phenomena, hinders efficient population inversion and compose a serious issue to achieve lasing in the standard (non-excitonic) picture at moderate particle densities[6].

Fortunately, when embedded in high quality factor micro-cavities, the coupling of strongly absorbing and emissive excitonic media can yield the formation of exciton-polaritons, which are composite bosons emerging from strong light-matter coupling conditions[8]. As opposed to uncoupled cavity photons, cavity polaritons do interact efficiently with their environment, and the bosonic final state stimulation of their scattering has been identified as an efficient mechanism for the formation of spatially and temporally coherent states without the stringent requirement of population inversion[9].

While at thermal equilibrium, such a phenomenon is commonly described in the framework of a Bose-Einstein condensate[10], the kinetic nature of polariton quantum liquids, arising from rapid particle tunnelling out of the microcavity puts the bare formation of coherent states in the polariton laser class[11]. Such devices, thus far, were realised in GaAs and II/VI microcavities at cryogenic temperatures[11,12] and under electrical injection[13], and realized in GaN[14], and later in organic[15] and perovskite[16] microcavities at ambient conditions.

Our work tackles the crucial question, whether exciton-polaritons, which are created in an atomically thin crystal sheet of $WSe_2$[17–20] coupled to an optical microcavity, can emit coherent light at room temperature. Indeed, utilizing atomically thin transition metal dichalcogenides in polaritonic light sources and condensates can pave the way towards a multiplicity of interesting applications and experiments. The locking of spin and valley in those materials profoundly changes optical selection rules, and the resulting chirality can be harnessed in topological nanophotonic applications[21,22]. The possibility of hosting excitons and free carriers of substantial density in one, or multiple monolayers holds big promise towards the study of high-density Bose-Fermi mixtures in polaritonics. Finally, the possibility to expand towards multiple, even rotationally aligned monolayers facilitates to harness *twistronics* approaches[23] to engineer novel quantum states of light.

In our experiment, we scrutinize density dependent phenomena in the radiation from exciton –polaritons based on a single $WSe_2$ crystal in a microcavity. While we capture signatures of a polaritonic threshold already at modest pump-powers, the interacting nature of polaritons becomes particularly evident at high pumping conditions where a frequency blueshift dependent on the free carrier density in the monolayer is observed. The valley character of our polaritons is evident in magneto-optical measurements, revealing a substantial valley-Zeeman effect. Most importantly, we find a clear onset of spatial coherence in our confined polariton modes, which is retained for a characteristic timescale of 4.5 ps.

**RESULTS**

**Sample and polariton dispersion relation**

Figure 1a shows a scheme of the sample structure. The bottom distributed Bragg reflector (DBR) is composed of 10 pairs of $SiO_2/TiO_2$ films, and its central Bragg wavelength is 750 nm. The embedded heterostructure consists of a $WSe_2$ monolayer covered by a h-BN layer, this set is assembled via the deterministic dry-transfer method. The top DBR is deposited via a carefully adapted evaporation routine[24] and consists of 9 pairs of $SiO_2/TiO_2$ (see Fig. 1a). We measure a quality factor of 4300±400 (see Fig. S9).

Similar to previous studies on analogous devices, which are detailed in Ref. [25], the $WSe_2$ monolayer under study in this work has a finite size of approximately 10 x 7 µm² (see dashed area in Fig. 1b). The monolayer flake is surrounded by few-layer areas as well as the empty cavity regions that can act as a trap. The geometry of the monolayer area defines the size of the polariton trap.

Throughout this work, we study the sample under ambient conditions with a standard confocal setup, and utilize a continuous wave green laser (setup details are described in the Methods section), focused to a spot of 3 μm diameter.

The angle-resolved PL spectrum of our sample is shown in Fig. 1c. The spectrum is recorded under moderate pumping conditions. It is composed of signals coming from the lower branch of the exciton-polariton dispersion at finite momenta (>2 μm$^{-1}$), and more significantly, it adds the features of a polariton trap at low energies: the ground state is a dispersion-less, discrete mode at 1.610 eV with a linewidth of 1.7 meV. This mode is separated from a ladder of polariton eigenstates arising from the reduced and irregular size of the flake.

To gain a deeper understanding of the mode dispersion, in Fig. 1d, we calculate the dispersion relation of excitons-polaritons by solving the Schrödinger equation for excitons-polaritons in a trap based on realistic system parameters: The energy of the WSe$_2$ monolayer exciton was experimentally determined as 1.67 eV, and the energy of the cavity photon is 1.612 eV, as confirmed by the transfer matrix calculations. The free parameter of our model, thus, is the Rabi-energy, which we extract from the fit to the experimental data as ~30 meV. The size of the optical trap is 10 x 7 μm$^2$, yielding the discretised ground state. By utilizing these parameters, our theory reproduces all experimental features of the ladder-like polariton dispersion and allows us to conclude on the potential depth of ~15 meV (see supplementary material section S1 for further details on the calculation of the polariton dispersion relation).

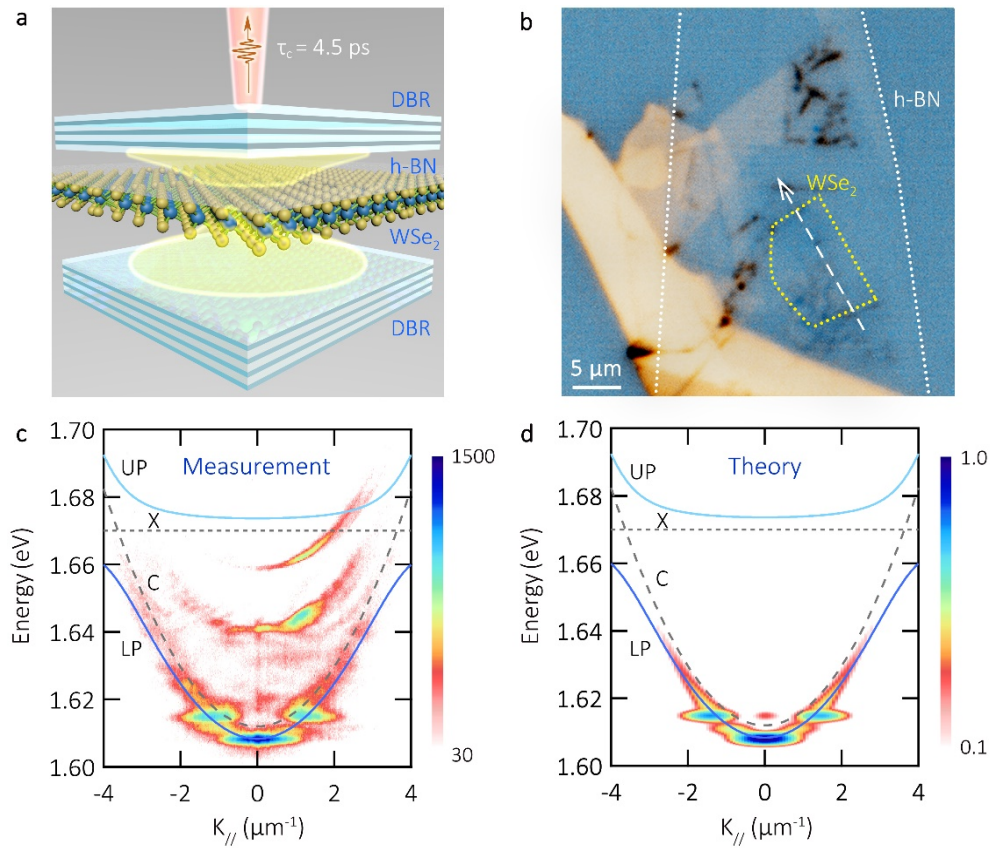

**Fig. 1 Sample sketch and polariton dispersion relation. a** Schematic of sample structure composed of two DBRs and an encapsulated monolayer of WSe$_2$. We schematically represent the emission from the sample displaying a coherence time of 4.5 ps. **b** Optical microscope image of the sample. The WSe$_2$ monolayer (h-BN) is highlighted with yellow (white) dash lines. A white dashed arrow indicates the slice resolved in the energy versus real space measures, the pump spot is placed in the centre of the monolayer. **c** Room temperature polariton dispersion relation encoded in a false logarithmic colour-

scale. The pump power is 3 µW. The dotted and dashed lines represent the bare exciton and microcavity modes, respectively. The resulting upper (UP) and lower polariton (LP) dispersion relations are shown as guides to the eye in light and dark blue colours, respectively. The exciton and photon modes are labelled X and C, respectively. The localised and bright emission modes at low energy arise from trapped polaritons. **d** Corresponding simulated dispersion distribution of exciton-polaritons. The energies are obtained via solving Schrödinger's equation for polaritons in a trap. The intensities are obtained assuming a thermal distribution of particles.

**Magnetic response of room temperature polaritons**

Excitons in $WSe_2$ monolayers can be formed in the two energetically degenerate K and K' valleys which are locked to opposite spins. In the presence of an externally applied magnetic field, the valley degeneracy is lifted via the valley-Zeeman effect[26–30] resulting in two circularly polarized optical transitions. Consequently, exciton-polaritons formed by valley excitons display an analogous Zeeman energy-splitting under a magnetic field[31], which clearly distinguishes them from cavity photons, and which provides a unique control knob in Valley Polaritonics[32].

In Figs. 2a,b, the top panels represent the PL intensity distribution map at room temperature under a magnetic field of -8/+8 T, respectively. The corresponding degree of circular polarization (DOCP) is presented in the bottom panels. The confined polariton mode (see dashed grey rectangle) shows a strong valley dependence and the spectrally resolved DOCP is inverted as the magnetic field changes from negative to positive (additional data is shown in Figs. S11, S12 in the Supplementary Material, no DOCP is observed at zero magnetic field). Analysing the trapped state within the grey box, in Fig. 2c we present the normalized PL intensities of σ⁺ and σ⁻ emission at -8T (top) and 8T (bottom panel), respectively. They clearly show an energy splitting between σ⁺ and σ⁻ emission.

This Zeeman energy splitting in the trapped state is systematically studied as a function of the applied magnetic field in Fig. 2d. A line with a slope of 0.081±0.004 meV/T fits the experimental data points. From this result, the Zeeman energy splitting can be linked to the polariton $g$-factor ($g_{pol}$), following the relation $E_{\sigma^+} - E_{\sigma^-} = g_{pol}\mu_B B$, where $\mu_B$ is the Bohr magneton and $B$ is the magnetic field. Provided the fitted slope of Zeeman splitting vs. magnetic field reported in Fig. 2d, we obtain $g_{pol} = 1.40 \pm 0.07$.

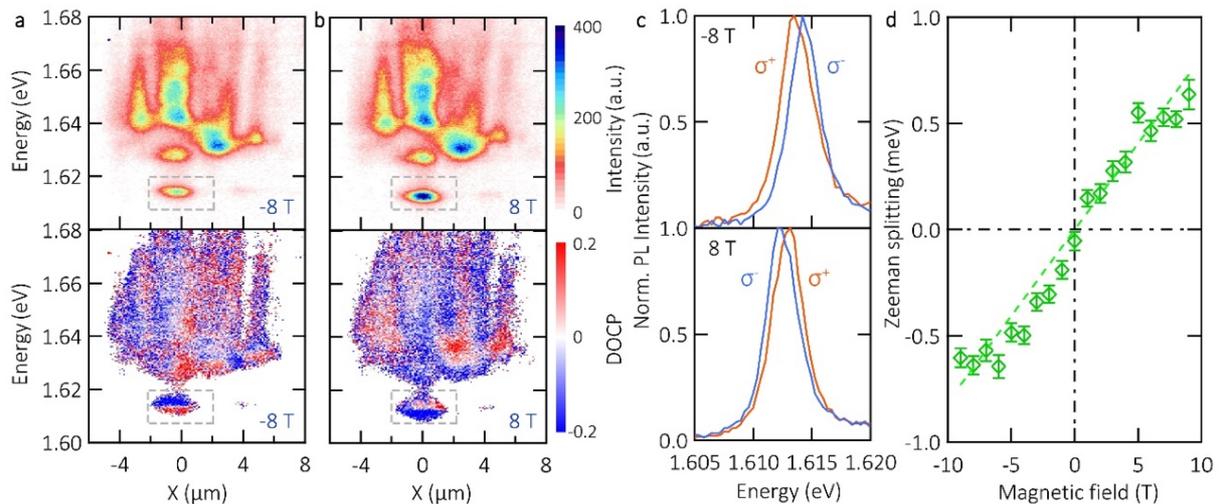

**Fig. 2 Magnetic response of room temperature exciton-polaritons. a, b** Polariton PL distribution (Iσ⁺+Iσ⁻) (top panel) and corresponding DOCP (bottom panel) as a function of energy and real space at a magnetic field of -8 T and 8 T, respectively. The pump power is 100 µW. Points in the (Iσ⁺+Iσ⁻) panels with <10% of the maximum intensity, are encoded in with white colour in the bottom panels. A dashed grey box in these panels highlights the region of interest in this Zeeman study, corresponding to the trapped polariton state. **c** Normalized PL intensity of σ⁺ (orange) and σ⁻ (blue) emission at magnetic field of -8 T (top) and 8 T (down). **d** Zeeman splitting as a function of the magnetic field. The dashed line is a linear fit with a slope of 0.081±0.004 meV/T.

**Density dependent properties of room temperature polaritons**

To assess the fundamental features of our sample throughout the regimes from low to high quasiparticle densities, we study the angle-resolved PL emitted from the lowest energy (trapped) mode as a function of the pump power. In Figs. 3a-c we represent three exemplary dispersion relations at 0.7, 5, and 90 µW, respectively, in three different points of the polariton input-output curve (the acquisition time is 180 s for each pump power). Following the counts in the false-color scale, we observe an increase of polariton PL, predominantly emitting from the trapped state at ~1.61 eV. We integrate the counts around this energy state and we represent its averaged intensity versus pump power in Fig. 3d; the red arrows in this panel indicate the corresponding dispersion relations plotted in panels a-c.

The input-output curve clearly displays deviations from the linear trend and features a kink at a pump power of ~1 µW, as well as the reduction of the slope towards larger pump densities, resembling the soft shape of an S. Since our device is operated under strong coupling conditions, the population of the trapped mode needs to be treated in a kinetic model based on the Boltzmann equations, where we consider phonon- and polariton-polariton scattering channels (see Supplementary Material section S2 for further details on the model), as well as the experimentally extracted parameters of our sample. The result of this model is plotted alongside the experimental data, shown as the solid line in Fig. 3d. It corroborates our findings of a smooth S-curve induced by quasiparticle scattering to the final state. We attribute this input-output line shape to a signature of polariton condensation threshold.

Simultaneously, in Fig. 3e we investigate the blueshift (blue, left axis) and linewidth (orange, right axis) of the trapped state versus pump power. The blueshift increases slowly up to a pump power of 40 µW, while a stronger increase up to 0.4 meV is observed in the high-density regime.

We argue that the dominant contribution of this blueshift arises from a fermionic screening of excitons by free carriers, which consequently quenches the Rabi-splitting in our system. The accumulation of free carriers in TMDC monolayers subject to optical pumping is a well-documented phenomenon, both at cryogenic and ambient conditions. It occurs over multiple minutes and hours, and hence during the full measurement of the input-output curve[33]. We emphasize, that the resulting energy renormalization, arising from the screening effect, is a clear sign of strong coupling conditions[34,35], which cannot be explained for a cavity in the weak coupling regime. Interestingly, we reconfirm the nature of the charge-induced blueshift by repeating the input-output experiment via reducing the pump power, yielding an 'inverted' blueshift of the mode (see supplementary section S3, Fig. S1). The linewidth of the trapped state, which is an indicator for temporal coherence of the system, only features a mild decrease within the measurement accuracy, which calls for a deeper analysis utilizing more sensitive interferometric tools, as we provide later in this work.

In Figs. 3f,g we investigate the linear polarization properties of the trapped state versus pump power. For this study, we place a half-waveplate and a linear polarizer in the collection path, providing the linear polarization dependence of the trapped state as a function of the half-waveplate angle. Figure 3f shows the average intensity of the trapped state (the studied area is indicated in a dashed box in panel c) analyzing its linear polarization, for three pump powers, 5, 30 and 85 µW. The full rotation of the half-waveplate results in four periods of intensity oscillation, evidencing the presence of linearly polarized emission from the trapped state.

We extract the corresponding visibility of the oscillations, retrieving the degree of linear polarization (DOLP) for the three different pump powers: 9±2%, 12±2%, 17±1%. Overall, we find clear enhancement of the DOLP as a function of pump power. This linear polarization predominance in the trapped state is expected from the polariton macroscopic occupation state under strong excitation.

The trapped state presents two orthogonally-polarized modes, H and V. Fig. 3g represents the spectra of these modes, revealing a fine structure splitting of 0.40±0.03 meV. The linear polarisation orientation of the pump does not affect the resulting DOLP of the trapped state (Fig. S3).

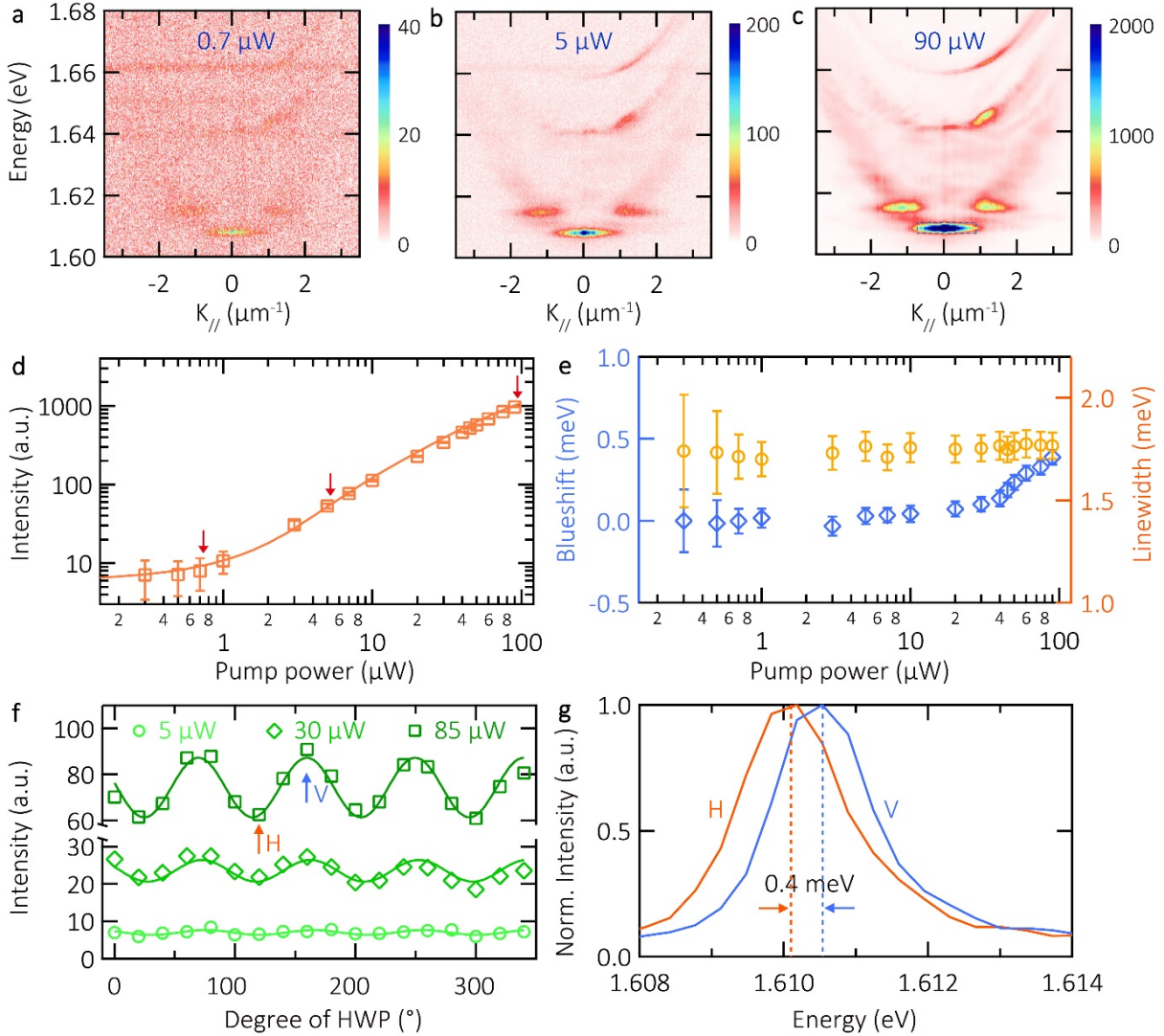

**Fig. 3 Density-dependent properties of room temperature exciton-polaritons. a-c** Polariton dispersion relations under 0.7, 5 and 90 µW pump power excitation. Intensity counts detected in the CCD are encoded in a false color scale. The acquisition time for each dispersion relation is 180 s. **d** Integrated PL emission intensity as a function of pump power, plotted in a double-logarithmic scale. The solid line is a simulation obtained by solving the Boltzmann equation. The arrows indicate the corresponding dispersion relations shown in panels a-c. The error bars are obtained by comparing the polariton signal intensity to the standard deviation of the background noise. **e** Corresponding blueshift (blue diamonds) and linewidth (orange circles) of exciton-polaritons as a function of pump power. The blueshift and linewidth error bars correspond to the 95% confidence interval of the peak fitting. **f** Linear polarization emission intensity analyzed as a function of the half-waveplate angle at three different pump powers. Each data point corresponds to the averaged intensity within the trapped state. The labels H and V indicate the two orthogonally-polarized states of the trap. **g** Corresponding H and V spectra, displaying the trap fine structure splitting of 0.40±0.03 meV.

## Coherence of room temperature polaritons

In this section, we present the build-up of macroscopic phase coherence in the polariton emission. The degree of spatial and temporal coherence is investigated via interferometric experiments, where the first-order correlation function $G^{(1)}(r,t)$ is experimentally extracted from measurements in a Michelson interferometer (see the sketch of the interferometer in Fig. 4a).

Due to the cavity mode emission at higher energies (see dispersion relation in Fig. 1c, cavity modes present at ~1.64 and ~1.66 eV), and the energy discretized polariton modes, we focus our coherence analysis in a one-dimensional slice centered in the sample (see dashed white arrow in Fig. 1b). The polariton emission is interfered with a retroreflected version of itself in a Michelson interferometer. We image the real space distribution of the interference and we spectrally resolve the marked slice in the spectrometer. In this way, we investigate the energy-resolved coherence of the polariton emission in a one-dimensional slice of the real space emission.

A representative set of images from the interferometric measurement is displayed in Figs. 4b-d. They correspond to the reference arm, the interference at zero temporal delay, and the resulting Fourier-extracted $G^{(1)}$ map, respectively, encoded in a false-color scale (see Methods and Supplementary Material for further details on the interference measurements). The pump power used in this measurement is 90 µW. The PL from the reference arm shows three dominant set of modes: the polariton ground state (~1.610 eV), a second discrete mode around X=0, as well as a two-lobed mode evolving slightly above ~1.615 eV. This set of modes can be retrieved and clearly identified via tomographic real-space scans of the trap (see Fig. S4). Above the energy 1.62 eV, we detect weak emission features that we attribute to high energy trapped states. One of the interference arm images is retroreflected, generating a vertical and horizontal inversion of the image. The interference map is obtained with the previous image and the delay arm. We extract the corresponding $G^{(1)}$ map from the Fourier transform of the interference pattern, and the delay and reference arms. Our analysis reveals a high degree of coherence in the discrete polariton modes localised below 1.62 eV. The peak of the $G^{(1)}$ map is localised in the lowest energy mode, reaching a value of ~0.25.

Next, we investigate the temporal coherence length of the polariton emission at this same pump power (90 µW). This measurement is done by displacing the delay arm of the interferometer, and recording the resulting interferences for different relative delays between the two interferometric arms. In our analysis, we focus on two different regions I and II indicated in dashed boxes in Fig. 4d. We average the $G^{(1)}$ values in these two regions, yielding the $<G^{(1)}>$ dependence in each region versus delay, see Fig. 4e. From a Gaussian fit to the experimental data points, we extract a coherence time of 4.5 (3.6) ps FWHM in region I (II), which reflects the monochromatic nature of the modes of interest in our study.

Lastly, we investigate the coherence buildup as a function of the polariton density. Following the same analysis procedure as that reported in Fig. 4e, in panel f we fix the interferometer delay at zero and we study the $<G^{(1)}>$ values for different pump powers of excitation (see Supplementary Materials for the full maps of interference and $G^{(1)}$ as a function of pump power). In both regions I and II we observe an increase and saturation of $<G^{(1)}>$ at high pump power excitation, demonstrating the phase-locking of a collective polariton state at room temperature. This feature clearly distinguishes the polariton condensate from incoherent polaritons in the linear regime. In Fig. 4g we study the coherence length $\lambda_c$ as a function of pump power in region I. We fit a Gaussian function to the spatial profile of $G^{(1)}$ in the trapped state (region I), and we associate the corresponding coherence length to the standard deviation σ of the fit: $\lambda_c = (2\pi)^{1/2}\sigma$. It is interesting to see that at low pump powers, $\lambda_c$ is comparable to the laser spot diameter, and at high pump powers, it reaches approximately the size of the polariton trap (see Fig. S8 for further details on the coherence length analysis of the different states in Fig. 4d).

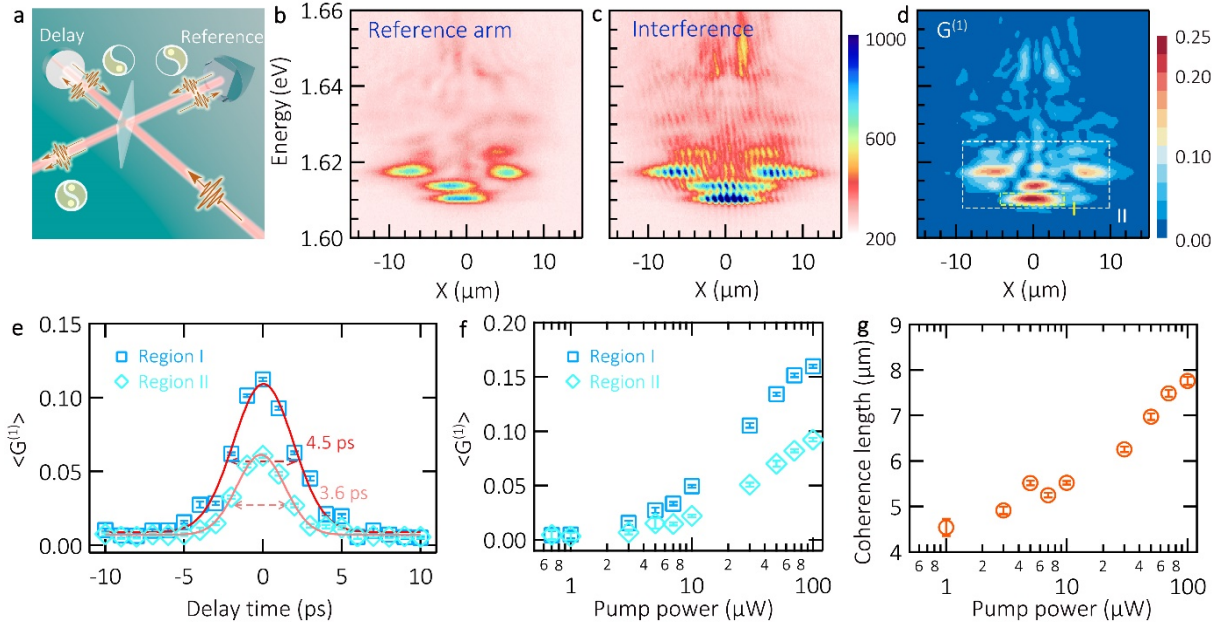

**Fig. 4 First order autocorrelation measurement of room temperature exciton-polaritons. a** Sketch of the Michelson interferometer, containing a retroreflector in the reference arm (fixed position, providing a vertical and horizontal image mirroring) and a flat mirror in the delay arm (mounted on a motorized translation stage). The "yin and yang" symbols represent, schematically, the two interfering images. **b** PL distribution of polaritons as a function of energy and real space (see the resolved X slice in Fig. 1b), recorded from the reference arm of Michelson interferometer. **c** Corresponding interference image at zero delay time between reference and delay arms. Intensity is encoded in a false color scale. The acquisition time for each image is 120 s. **d** Corresponding first order correlation function $G^{(1)}$ encoded in a false color scale. **e** Delay-time dependence of $<G^{(1)}>$, the data points are the result of averaging $G^{(1)}$ in regions I and II (dashed boxes in panel d), displayed as squares and diamonds, respectively. The data is fitted with a Gaussian function, indicating the corresponding temporal FWHM. The pump power is 90 μW in panels b-e. **f** $<G^{(1)}>$ as a function of pump power. The square (diamond) data points correspond to region I (II). In panels e and f the error bars are obtained by comparing the polariton $G^{(1)}$ signal to the standard deviation of the background $G^{(1)}$ noise. **g** Coherence length $\lambda_c$ in region I as a function of pump power, the error bars correspond to the 95% confidence interval of the Gaussian fitting from the $G^{(1)}(x)$ spatial shape of the polariton trap.

## DISCUSSION

Our density-dependent study of monolayer exciton-polaritons at ambient conditions allows us to conclude that the observed input-output characteristics of the studied structure strongly hint at the formation of a bosonic condensate of exciton-polaritons. The experimental data are in full agreement with a Boltzmann-model, which accounts for stimulated photon- and polariton-polariton scattering to the system ground state, which are the two essential prerequisites to obtain a macroscopic population in the polaritonic ground state. Our data suggest, that the overall energy renormalization of the trapped polariton mode is dominated by free-carrier screening, which clearly reflects the polaritonic nature of our system. Furthermore, this effect is highly promising for future experiments to scrutinize the complex interplay of bosons and fermions in macroscopic coherent states. The characteristic features of polariton lasing, such as the built-up of linear polarization, and in particular as the spatial and temporal coherence induced by the increase of quasi-particle densities are all observed in our system. These observations certify the realization of the first generation of room temperature coherent light emitters based on atomically thin crystals. We highlight, that our room temperature polariton condensates display clear features associated with the intricate valley physics provided by the TMDC monolayer. In our particular case, we scrutinize the valley Zeeman-effect, which enables to tune the polarization and energy structure of our TMD polaritons, and highlights the promise of coherent polaritonic TMDC systems in nanophotonic valleytronic devices. We recently became aware of similar results to the ones reported in this paper, based on WS$_2$-based excitons[36].

## METHODS

The excitation path contains two different light sources: a 532 nm CW laser, and a white light source (Thorlabs SLS301). The sample is mounted in a motorized XYZ stage, model Attocube ECSxy5050/Al/NUM/RT and ECSz5050/Al/NUM/RT. The magnification of the real space imaging is x110 (the first lens on top of the sample is a Thorlabs C105TMD-B). The optical setup uses a standard back Fourier plane imaging with a magnification of x1.3. The polarisation control at the excitation and detection (set of linear polariser, half- and quarter-waveplate) are calibrated with a polarization analyser Schäfter+Kirchhoff SK010PA. The polarisation excitation (detection) set is placed immediately before (after) the input (output) mode of the beam splitter, in reflection (transmission).

The measures are performed in a CCD camera, model iKon-M DU934P-BEX2-DD-9FL, attached to a spectrometer, model A-SR-500i-B2-SIL. The CCD camera is operated in the regime for highest sensitivity (50 kHz A/D rate, pre-amplifier setting x4, and sensor cooling temperature of -80°C). In the collection path, we include a compact Michelson interferometer composed by a 50:50 beam splitter (Thorlabs BSW11R), a retroreflector (Thorlabs PS976M-B) in the reference arm and a flat silver mirror in the delay arm, mounted in a translation motorized stage (with a motor Thorlabs Z825B). The zero-delay calibration of the interferometer is obtained using 3 ps laser pulses. The experimental splitting ratios of the Michelson interferometer are 0.54:0.46.

The magnetic field measurements are performed with a cryostat attoDRY2100 operated at room temperature. A linearly polarised, 532 nm CW laser excites the sample. A set of quarter-half-waveplates and linear polarizer are used to detect the $\sigma^+$ and $\sigma^-$ polarized components of the emission. Due to the low brightness of the sample, we perform these experiments recording the real space distribution of the emission, which provides a more localised PL signal than the momentum space distribution. The DOCP is calculated as $(I_{\sigma+} - I_{\sigma-})/(I_{\sigma+} + I_{\sigma-})$ where $I_{\sigma+}$ and $I_{\sigma-}$ are PL intensities of right- and left-hand circularly polarized emission, respectively. It is notable to remark that the real space PL distribution in Figs. 2a,b is different from Fig. 4b. In these magnetic field experiments, the sample is loaded in a cryo-system with a different orientation with respect to the spectrometer slit.

## DATA AVAILABILITY

The data that support the findings of this study are available from the corresponding author upon reasonable request.

## References

1. Mak, K. F. & Shan, J. Photonics and optoelectronics of 2D semiconductor transition metal dichalcogenides. *Nature Photon* **10**, 216–226 (2016).

2. Sun, Z., Martinez, A. & Wang, F. Optical modulators with 2D layered materials. *Nature Photon* **10**, 227–238 (2016).

3. Xia, F., Wang, H., Xiao, D., Dubey, M. & Ramasubramaniam, A. Two-dimensional material nanophotonics. *Nature Photonics* **8**, 899–907 (2014).

4. Wang, Q. H., Kalantar-Zadeh, K., Kis, A., Coleman, J. N. & Strano, M. S. Electronics and optoelectronics of two-dimensional transition metal dichalcogenides. *Nature Nanotech* **7**, 699–712 (2012).

5. Yin, Z. *et al.* Single-Layer $MoS_2$ Phototransistors. *ACS Nano* **6**, 74–80 (2012).

6. Lohof, F. *et al.* Prospects and Limitations of Transition Metal Dichalcogenide Laser Gain Materials. *Nano Lett.* **19**, 210–217 (2019).

7. Paik, E. Y. *et al.* Interlayer Exciton Laser with Extended Spatial Coherence in an Atomically-Thin Heterostructure. *Nature* **576**, 80–84 (2019).

8. Weisbuch, C., Nishioka, M., Ishikawa, A. & Arakawa, Y. Observation of the coupled exciton-photon mode splitting in a semiconductor quantum microcavity. *Physical Review Letters* **69**, 3314–3317 (1992).

9. Imamoglu, A., Ram, R. J., Pau, S. & Yamamoto, Y. Nonequilibrium condensates and lasers without inversion: Exciton-polariton lasers. *Physical Review A* **53**, 4250–4253 (1996).

10. Kasprzak, J. *et al.* Bose–Einstein condensation of exciton polaritons. *Nature* **443**, 409–414 (2006).

11. Deng, H., Weihs, G., Snoke, D., Bloch, J. & Yamamoto, Y. Polariton lasing vs. photon lasing in a semiconductor microcavity. *P.N.A.S.* **100**, 15318–15323 (2003).

12. Dang, L. S., Heger, D., Andre, R., Boeuf, F. & Romestain, R. Stimulation of Polariton Photoluminescence in Semiconductor Microcavity. *Physical Review Letters* **81**, 3920–3923 (1998).

13. Schneider, C. *et al.* An electrically pumped polariton laser. *Nature* **497**, 348–352 (2013).

14. Christopoulos, S. *et al.* Room-Temperature Polariton Lasing in Semiconductor Microcavities. *Phys. Rev. Lett.* **98**, 126405 (2007).

15. Kena Cohen, S. & Forrest, S. R. Room-temperature polariton lasing in an organic single-crystal microcavity. *Nature Photonics* **4**, 371–375 (2010).

16. Su, R. *et al.* Room-Temperature Polariton Lasing in All-Inorganic Perovskite Nanoplatelets. *Nano Lett.* **17**, 3982–3988 (2017).

17. Dufferwiel, S. *et al.* Exciton–polaritons in van der Waals heterostructures embedded in tunable microcavities. *Nat Commun* **6**, 8579 (2015).

18. Liu, X. *et al.* Strong light–matter coupling in two-dimensional atomic crystals. *Nature Photon* **9**, 30–34 (2015).

19. Lundt, N. *et al.* Room-temperature Tamm-plasmon exciton-polaritons with a WSe2 monolayer. *Nat Commun* **7**, 13328 (2016).

20. Schneider, C., Glazov, M. M., Korn, T., Höfling, S. & Urbaszek, B. Two-dimensional semiconductors in the regime of strong light-matter coupling. *Nat Commun* **9**, 2695 (2018).

21. Liu, W. *et al.* Generation of helical topological exciton-polaritons. *Science* eabc4975 (2020) doi:10.1126/science.abc4975.

22. Lundt, N. *et al.* Optical valley Hall effect for highly valley-coherent exciton-polaritons in an atomically thin semiconductor. *Nat. Nanotechnol.* **14**, 770–775 (2019).

23. Kennes, D. M. *et al.* Moiré heterostructures as a condensed-matter quantum simulator. *Nat. Phys.* **17**, 155–163 (2021).

24. Knopf, H. *et al.* Integration of atomically thin layers of transition metal dichalcogenides into high-Q, monolithic Bragg-cavities: an experimental platform for the enhancement of the optical interaction in 2D-materials. *Opt. Mater. Express* **9**, 598 (2019).

25. Rupprecht, C. *et al.* Demonstration of a polariton step potential by local variation of light-matter coupling in a van-der-Waals heterostructure. *Opt. Express* **28**, 18649 (2020).

26. Srivastava, A. *et al.* Valley Zeeman effect in elementary optical excitations of monolayer WSe2. *Nature Phys* **11**, 141–147 (2015).

27. Aivazian, G. *et al.* Magnetic control of valley pseudospin in monolayer WSe2. *Nature Phys* **11**, 148–152 (2015).

28. Stier, A. V. *et al.* Magnetooptics of Exciton Rydberg States in a Monolayer Semiconductor. *Phys. Rev. Lett.* **120**, 057405 (2018).

29. Stier, A. V., Wilson, N. P., Clark, G., Xu, X. & Crooker, S. A. Probing the Influence of Dielectric Environment on Excitons in Monolayer $WSe_2$ : Insight from High Magnetic Fields. *Nano Lett.* **16**, 7054–7060 (2016).

30. Molas, M. R. *et al.* Energy Spectrum of Two-Dimensional Excitons in a Nonuniform Dielectric Medium. *Phys. Rev. Lett.* **123**, 136801 (2019).

31. Anton-Solanas, C. *et al.* Bosonic condensation of exciton–polaritons in an atomically thin crystal. *Nat. Mater.* (2021) doi:10.1038/s41563-021-01000-8.

32. Rupprecht, C. *et al.* Manipulation of room-temperature valley-coherent exciton-polaritons in atomically thin crystals by real and artificial magnetic fields. *2D Mater.* **7**, 035025 (2020).

33. Lundt, N. *et al.* The interplay between excitons and trions in a monolayer of $MoSe_2$. *Appl. Phys. Lett.* **112**, 031107 (2018).

34. Sidler, M. *et al.* Fermi polaron-polaritons in charge-tunable atomically thin semiconductors. *Nature Phys* **13**, 255–261 (2017).

35. Chakraborty, B. *et al.* Control of Strong Light–Matter Interaction in Monolayer $WS_2$ through Electric Field Gating. *Nano Lett.* **18**, 6455–6460 (2018).

36. Zhao, J. *et al.* Ultralow Threshold Polariton Condensate in a Monolayer Semiconductor Microcavity at Room Temperature. *Nano Lett.* **21**, 3331–3339 (2021).

**Acknowledgement**

The authors gratefully acknowledge funding by the State of Lower Saxony. Funding provided by the European Research Council (ERC project 679288, unlimit-2D) is acknowledged. ST acknowledges funding from NSF DMR 1955889, DMR 1933214, and 1904716. ST also acknowledges DOE-SC0020653, DMR 2111812, and ECCS 2052527 for material development and integration. AVK acknowledges Westlake University (Project No. 041020100118) and the Program 2018R01002 funded by Leading Innovative and Entrepreneur Team Introduction Program of Zhejiang. K.W. and T.T. acknowledge support from the Elemental Strategy Initiative conducted by the MEXT, Japan, Grant Number JPMXP0112101001, JSPS KAKENHI Grant Number JP20H00354 and the CREST(JPMJCR15F3), JST. HS acknowledges the Sino-Germany (CSC-DAAD) Postdoctoral Scholarship Program from China Scholarship Council and German Academic Exchange Service. AVK acknowledges the support from Rosatom within the Road map for Quantum computing programme. H.S., J.B. and C.A.-S. acknowledge Tobias Huber for his assistance in the magnetic field experiments.

**COMPETING INTERESTS**

The authors declare no competing interests.